# Segmented Terahertz Electron Accelerator and Manipulator (STEAM)

Dongfang Zhang[1,2*†], Arya Fallahi[1*], Michael Hemmer[1], Xiaojun Wu[1], Moein Fakhari[1,2], Yi Hua[1], Huseyin Cankaya[1], Anne-Laure Calendron[1], Luis E. Zapata[1], Nicholas H. Matlis[1], and Franz X. Kärtner[1,2,3]

**Affiliations:**

[1]Center for Free-Electron Laser Science, Deutsches Elektronen Synchrotron, Notkestrasse 85, 22607 Hamburg, Germany.

[2]Department of Physics and The Hamburg Center for Ultrafast Imaging, University of Hamburg, Luruper Chaussee 149, 22761 Hamburg, Germany.

[3]Research Laboratory of Electronics, MIT, Cambridge, 02139 Massachusetts, USA.

*Equal contribution.

†To whom correspondence should be addressed. E-mail: dongfang.zhang@cfel.de

**Abstract**: Acceleration and manipulation of ultrashort electron bunches are the basis behind electron and X-ray devices used for ultrafast, atomic-scale imaging and spectroscopy. Using laser-generated THz drivers enables intrinsic synchronization as well as dramatic gains in field strengths, field gradients and component compactness, leading to shorter electron bunches, higher spatio-temporal resolution and smaller infrastructures. We present a segmented THz electron accelerator and manipulator (STEAM) with extended interaction lengths capable of performing multiple high-field operations on the energy and phase-space of ultrashort bunches with moderate charge. With this single device, powered by few-microjoule, single-cycle, 0.3 THz pulses, we demonstrate record THz-device acceleration of >30 keV, streaking with <10 fs resolution, focusing with >2 kT/m strengths, compression to ~100 fs as well as real-time switching between these modes of operation. The STEAM device demonstrates the feasibility of future THz-based compact electron guns, accelerators, ultrafast electron diffractometers and Free-Electron Lasers with transformative impact.

## Introduction

Accelerator development over the last century has underpinned the study of material properties and structure at ever higher spatial and temporal resolutions. Until recently, microwaves in the radio-frequency (RF) regime (1-10 GHz) have been the conventional choice for powering accelerators due to the high degree of technical maturity of the sources which have been used extensively across all areas of industry and science, from cell phones, microwave ovens and radar to linear accelerators[1], bunch compressors[2,3] and high-resolution streak cameras[4]. The long driver wavelengths are ideal for accelerating electron bunches of high charge, it has become possible to generate ultrafast electron pulses with very high peak brightness and quality. However, RF-based accelerators require costly infrastructures of large size and power[5], limiting the availability of this key scientific resource. They also suffer from inherent difficulties in synchronization with lasers[6] which lead to timing drifts on the 100 femtosecond scale between the electrons, microwave drivers and optical probes, limiting the achievable temporal resolution.

Strong motivation thus exists for exploring alternative technologies that are compact, more accessible and adapted for pushing the resolution frontiers. Novel accelerator concepts thus primarily focus on laser-based approaches that provide intrinsic synchronization, allow scaling to smaller accelerator structures and can generate substantially stronger fields for acceleration and beam manipulation. These include dielectric laser accelerators (DLAs)[7], laser-plasma accelerators (LPAs)[8,9] and laser-based THz-driven accelerators[10,11], each with different advantages. A consequence of shifting to smaller scales is that less charge can be supported and creation of reliable structures can become more difficult. LPAs, for example, which boast extremely high acceleration gradients on the order of 100 GeV/m, generate acceleration structures dynamically and therefore suffer from instabilities, difficulties in controlling injection and low repetition rates. DLAs, which employ micron-scale structures, require extreme tolerances on alignment and control, and are limited to bunch charges in the sub-femtocoulomb range. THz-based accelerators, however, exist at an intermediate, millimeter scale that allows traditional fabrication techniques and supports moderate charge while still benefiting from compactness, low costs and strong driving fields. This balance makes THz-based acceleration an extremely promising technology for future devices.

So far, the development of THz-based accelerators has been limited by the lack of sufficiently energetic THz sources, but recent progress in efficient laser-based methods[12-14] has enabled generation of high-power, GV/m THz fields, opening new possibilities and spurring interest in THz-accelerator-related technologies. Proof-of-principle demonstrations include electron emission[15,16] and acceleration[10,11,17-21] as well as compression and streaking[22,23]. These experiments, although limited in charge, beam quality, energy gain and energy spread, have set the stage for development of practical, compact THz-based devices which can support sufficient charge and field gradients to realistically be used to boost performance of existing accelerators or as components of future compact accelerators and X-ray sources. In this letter, we demonstrate the first such device based on a layered, transversely pumped, waveguide structure. This segmented THz electron accelerator and manipulator (STEAM) device can dynamically switch between *accelerating, streaking, focusing and compressing* modes, can support multiple picocoulombs of charge and features intrinsic synchronization. Using only a few microjoules of single-cycle THz radiation, we demonstrate 50 MV/m average acceleration gradients, 2 kT/m focusing gradients (which are an order of magnitude beyond current electromagnetic lenses and comparable to active plasma lenses), the highest reported streaking gradient of 140 µrad/fs (making it well-suited for characterization of ultrafast electron diffractometer bunches down to 10 fs) as well as compression to ~100 fs. By increasing THz pulse energies to state-of-the-art millijoule levels[13], it is expected that acceleration gradients approaching GeV/m can be achieved, with similar gains in other modes of operation. These figures confirm the practicality of STEAM devices for future accelerators.

**Concept and implementation**

The experimental setup (Fig.1A) consisted of a 55 keV photo-triggered DC gun, a THz-powered STEAM device and a diagnostic section which included a second STEAM device, all of which were driven by the same IR laser source. Ultraviolet pulses for the photoemission were generated by two successive stages of second harmonic generation (SHG), while single-cycle THz pulses were generated by intra-band difference frequency generation. THz from two independent setups was coupled into the STEAM device (Fig. 1C) transversely to the electron motion by two trapezoidally-shaped horn structures which concentrated the counter-propagating THz fields beyond the diffraction limit into the interaction zone where electrons were injected from the DC gun. According to the Lorentz force law, the electrons experience both the electric and magnetic fields of the THz pulses according to $\vec{F} = e(\vec{E} + \vec{v} \times \vec{B})$, where $e$ is the electron charge, $\vec{E}$ is the electric field, oriented parallel to the electron velocity $\vec{v}$, and $\vec{B}$ is the magnetic field, oriented vertically in the lab frame. The electric field is thus responsible for acceleration and deceleration, while the magnetic field induces transverse deflections.

Efficient interaction of the electrons with the fields was accomplished by means of segmentation which divided the interaction volume into multiple layers of varying thickness, each isolated from the others by thin metal sheets (Fig. 1A). Dielectric slabs of varying length were inserted into each layer to delay the arrival time of the THz waveform to coincide with the arrival of the electrons, effectively quasi-phase-matching the interaction. Due to the transverse geometry, the degree of dephasing experienced in each layer was determined by the traversal time of the electrons, which was dependent on the electron speed and the layer thickness. A reduction in dephasing can thus be accomplished by reducing the layer thickness and increasing the number of layers, at the cost of increased complexity. The ability to tune the thickness and delay of each layer independently is a key design feature of the STEAM device that enables acceleration of sub-relativistic electrons for which the speed changes significantly during the interaction (e.g, from 0.43 *c* to 0.51 *c* for our maximum acceleration case).

The use of two counter-propagating drive pulses enabled two key modes of operation: an "electric" mode, used for acceleration, compression & focusing, in which the fields were timed to produce electric-field superposition and magnetic-field cancellation; and a "magnetic" mode, used for deflection and streaking, where the magnetic fields superposed and the electric fields cancelled. The function of the device was thus selected by tuning the relative delay of the two THz pulses and the electrons, all of which were controlled by means of motorized stages acting on the respective IR pump beams. In focusing and streaking modes the electron beams were sent directly to a camera and micro-channel plate (MCP) detection system which recorded the beam spatial distributions. For acceleration measurements, an electromagnetic dipole was used to induce energy-dependent deflections in the vertical plane, so that both deflection and energy change could be measured simultaneously. To measure the compression, a second STEAM device in streaking mode was added downstream of the first to induce time-dependent deflections in the horizontal plane.

Experiments were performed both with a 4 mJ, 1030 nm Yb:KYW laser operating at 1 kHz, to demonstrate high repetition rates, as well as with a 40 mJ, 1020 nm Yb:YLF laser operating at 10 Hz, in order to demonstrate high peak accelerations. Using these systems, THz pulses with a center frequency of 0.3 THz were generated (Fig. 1B) by the well-established tilted pulse-front (TPF) method[12], resulting in 2×2 µJ pulses and 2×30 µJ from the Yb:KYW and Yb:YLF laser systems respectively. Substantial losses due in transport and coupling of the THz beam, however, resulted in pulse energies at the interaction region of only 2×1 µJ and 2×6 µJ for the Yb:KYW and Yb:YLF respectively. The STEAM device was designed with three layers of thickness $h = \{0.225, 0.225, 0.250\}$ mm and with dielectric slabs in the second and third layers made of fused silica ($\epsilon_\mathrm{r} = 4.41$) and of length $L = \{0.42, 0.84\}$ mm. The entrance and exit apertures were of diameter 120 µm. The remainder of this paper gives a detailed description of the results obtained for acceleration, compression, focusing, deflection and streaking for this three-layer STEAM device.

**Electric Mode: acceleration, compression & focusing**

In the electric mode, the relative timing of the THz pulses was adjusted so that the electric fields (*E*-fields) constructively interfered at the interaction point, optimizing the acceleration of electrons. In this configuration, the magnetic fields (*B*-fields) were 180° out of phase with each other and thus cancelled, minimizing unwanted deflections. The acceleration was sensitive to the THz phase at the interaction, which was controlled by the arrival time of the electrons. Figure 2 shows energy and deflection diagrams which were obtained by recording the vertical and horizontal projections (respectively) of the electron-beam distribution on the MCP as a function of the electron-THz delay. Although the THz pulses injected into the device were nearly single cycle, several cycles of acceleration and deceleration were observed, due to dispersion induced by the horn couplers, which resulted in chirped, few-cycle pulses at the interaction point.

Maximal acceleration and deceleration occurred at the crests of the THz fields (Fig. 2A) where the deflection was minimized (Fig. 2C) and the beam spatial distribution was also preserved (Fig. 2E left & right beams). The peak field inside the cavity is calculated to have reached ~700 kV/cm, based on comparisons of the measured THz energy transmitted through the device and the electron energy gain with simulation (described below). The energy gain scaled linearly with the applied field (Fig. 3B) and reached a record maximum of more than 30 keV (5 times larger than previous works[10]). In contrast to previous results showing simultaneous acceleration and deceleration, the energy spectrum can be seen to move cleanly to a higher energy (Fig. 3A), indicating that injected bunches were shorter than half the driver period. In fact, the bunches were measured (by the STEAM device in streaking mode) to have a duration of 670 fs, and thus occupied about 20% of the 3.33 ps period accelerating field. The increase in energy spread can thus be attributed in part to the variation of the *E*-field over the bunch temporal profile. The long duration of the injected bunch was a result of space-charge forces experienced during its travel from the DC gun. Use of a THz-based re-buncher before the accelerator is thus anticipated for future experiments to reduce energy spread. An additional, perhaps stronger contribution to the measured energy spread came from averaging the beam profile in the presence of more significant shot-to-shot variations in THz pulse energies from the Yb:YLF laser as compared to Yb:KYW laser.

The performance of the device was simulated using a finite-element based code[24]. Figure 3C shows snapshots of the electrons traversing the device and staying in phase with the field. Figure 3D shows the electron energy as a function of distance. The energy gain can be seen to occur in three uneven steps corresponding to the three layers. The unevenness and the presence of deceleration at some points are evidence of dephasing due to the fact that the structure was designed for higher THz energies. Simulations predict that MeV electron beams with up to 10 pC of charge are achievable by increasing the number of layers and extending THz pulse energies to the millijoule level[17] which is within the reach of current THz-generation methods[13].

At off-crest timings, the electrons experienced strong temporal gradients of the *E*-field resulting in large energy spreads (Fig. 2A). At the zero-crossing of the field, the gradient is maximized and the electrons see symmetric acceleration and deceleration but no net energy gain. In this mode, the *E*-field imparts a temporally-varying energy or "chirp" resulting in a velocity gradient that causes either compression or stretching (depending on the sign of the gradient) of the electron bunch as it propagates[25]. This technique, known as "velocity bunching" is an ideal application of THz technology, since the sub-mm-scale gradients allow bunch compression down to the femtosecond range. To test this concept, the applied THz energy was varied and a second STEAM device ("streaker") acting as a streak camera (described in the next section) was added to measure the bunch temporal profile at a point 200 mm downstream of the first device ("buncher").

Figure 4A shows the electron bunch temporal profiles measured at the streaker for various field strengths applied to the buncher. The initial decrease in bunch duration with increasing field confirms that the electrons arrive at the buncher with a space-charge-induced energy chirp. A minimum duration of ~100 fs FWHM was achieved after which the duration increases again (Fig. 4B), implying that for high fields the electrons temporally focus before the streaker and are over compressed by the time they are measured. The minimum bunch duration can thus be reduced by using stronger fields and a shorter propagation distance. As observed on the MCP-detector (Fig. 4B inset), a good e-beam profile is maintained during focusing. Figure 4C shows the evolution of the bunch duration with distance simulated for the minimum bunch duration case. The phase-space distributions in insets D, E & F show the reversal of the velocity correlation by the buncher and the eventual compression at the streaker location.

By placing the electrons at the zero crossing in the electric mode, the STEAM device can also operate as a focusing or defocusing element, as can be seen by the horizontal spreading of the beam profile in Figure 2C. This focusing effect is a consequence of the well-known Panofsky-Wenzel Theorem[26] which uses Gauss's law to show that longitudinal compressing and decompressing fields must be accompanied by transverse defocusing and focusing fields, respectively (illustrated in Figures 5C&D). The focusing was tested by monitoring the beam spatial profile at the MCP for varying THz pulse energies. Figure 5A shows the results for the focusing configuration, which corresponded to the longitudinal decompression condition. At best focus the electron beam diameter was reduced by 2 × compared to its input value. For higher field strengths, however, the device focal length became shorter than the 180 mm distance to the MCP, causing the measured beam size to increase again. Although in these experiments the MCP position was fixed, it can be expected, as in optics, that for shorter distances, significantly smaller focused beam sizes would be measured. The defocusing configuration is obtained by shifting the electron timing to the longitudinal compression condition, which occurs at an adjacent zero crossing of the opposite sign. In this case, the electron beam diameter increases monotonically with the THz field (Fig. 5 B), as expected. For both cases, the focusing performance is significantly beyond what is offered by conventional electrostatic[27] and proposed dielectric focusing structures[28] and is comparable to those of plasma lenses[29]. Peak focusing gradients of over 2 kT/m were calculated based on ~2×6 μJ of coupled THz energy. A small (less than a factor of 2) asymmetry is noticeable for the focusing strengths in the horizontal and vertical planes. This asymmetry is due to the asymmetry of interaction region which leads to stronger gradients in the vertical direction (Fig. 5C&D).

**Magnetic Mode: Deflection & streaking**

In the magnetic mode, the relative timing of the THz fields is different from that of the electric mode by a half period, resulting in reinforcement of the magnetic and cancellation of the electric fields. In this configuration, electron acceleration is minimized (Fig. 2B), and the *B*-field dominates the interaction causing a transverse deflection of the electron beam that depends on the THz phase at the interaction (Fig. 2D). When electrons are on crest, the deflection is maximized and the beam profile is also best preserved. In this mode, electron beams can be precisely steered (Fig. 2E top & bottom beams) by varying the THz pulse energy. Here, we achieved continuous control of the beam angle over a range of 70 mrad which was limited by the aperture of the device. Increasing the aperture would allow greater range but at the cost of a weaker deflection field, since the field confinement is affected.

Electrons at the zero crossing of the THz $B$-field, on the other hand, experience a deflection that is a very steep function of time enabling the temporal profiles of very short bunches to be resolved by projecting (or "streaking") them onto the spatial dimension of a detector. To test this concept, a first STEAM device was used in compression mode (as described above) to provide electrons of varying bunch durations at a second, downstream STEAM device which analyzed the temporal profiles by streaking. Figure 6A shows raw images of a temporally-long electron beam with the THz streaking field switched on and off. Streaking "deflectograms," generated by plotting a lineout of the spatial charge distribution along the streaking dimension as a function of delay relative to the THz field, are shown in Figure 6B for compressed and uncompressed electron bunches. The degree of streaking, indicated by the vertical extent of the deflectogram, depends clearly on the bunch duration and on the phase of THz field, as expected. For a THz energy of ~2×6 µJ coupled into the device, a maximum deflection rate of > 140 µrad/fs was achieved, corresponding to a temporal resolution below 10 fs. The resolution was limited here by the size of the unstreaked beam on the MCP of 350 µm, and thus could be improved by better focusing. These results represent the first use of THz $B$-fields for deflection and streaking as well as a new record in THz-based streaking gradient.

**Conclusions and outlook**

We have demonstrated a novel segmented THz electron accelerator and manipulator which provides high capabilities and a multiplicity of functions in a very compact device. The transverse pumping and segmented structure makes it possible to phase-match the electron-THz interaction for non-relativistic beams, making it ideal for use as a high-gradient photogun. The independent control of the counter-propagating THz pulse timings gives the STEAM device the ability to switch dynamically between acceleration, compression, focusing, deflection and streaking modes. The use of THz pulses also brings other advantages, including negligible heat loads, high repetition rates and compactness while still supporting substantial charge.

Using only 2×6 µJ of THz energy the STEAM device has demonstrated acceleration gradients of 50 MV/m (average), compression of a bunch from over 1ps to 100 fs, focusing strength of up to 2 kT/m and streaking gradients in excess of 140 µrad/fs, leading to a temporal resolution below 10 fs. By scaling to millijoule-level THz energies which are already available, the field strengths in the device can be increased by over an order of magnitude, leading to capabilities that far exceed those of conventional RF devices. The exceptional performance and compactness of this THz-based device makes it very attractive for pursuing electron sources, like ultrafast electron diffractometers, that operate in the few- and sub-fs range necessary for probing the fastest material dynamics[30,31]. In the pursuit of these sources, the demand is increasing for compact, high-gradient diagnostics and beam manipulation devices for novel and conventional accelerator platforms alike. In a large-scale facility like the Swiss FEL or the European XFEL, STEAM devices can be used to add new, powerful and adaptable capabilities without a major and costly restructuring of the machine. Potentially more significant are the advantages in terms of cost and accessibility that would come from using STEAM devices as the core components of an all-THz-powered compact, high-gradient accelerator with the ability to produce high-quality, controllable bunches of femtosecond or attosecond duration on a table top. The results here take the first step in demonstrating the feasibility of that vision.

**Acknowledgments:** Besides DESY and the Helmholtz Association, this work has been supported by the European Research Council under the European Union's Seventh Framework Programme (FP7/2007-2013) through the Synergy Grant AXSIS (609920) and the excellence cluster "The Hamburg Center for Ultrafast Imaging – Structure, Dynamics and Control of Matter at the Atomic Scale" (CUI, DFG-EXC1074), the priority program QUTIF (SPP1840 SOLSTICE) of the Deutsche Forschungsgemeinschaft and the accelerator on a chip program (ACHIP) funded by the Betty and Gordon Moore foundation.

**Figures**

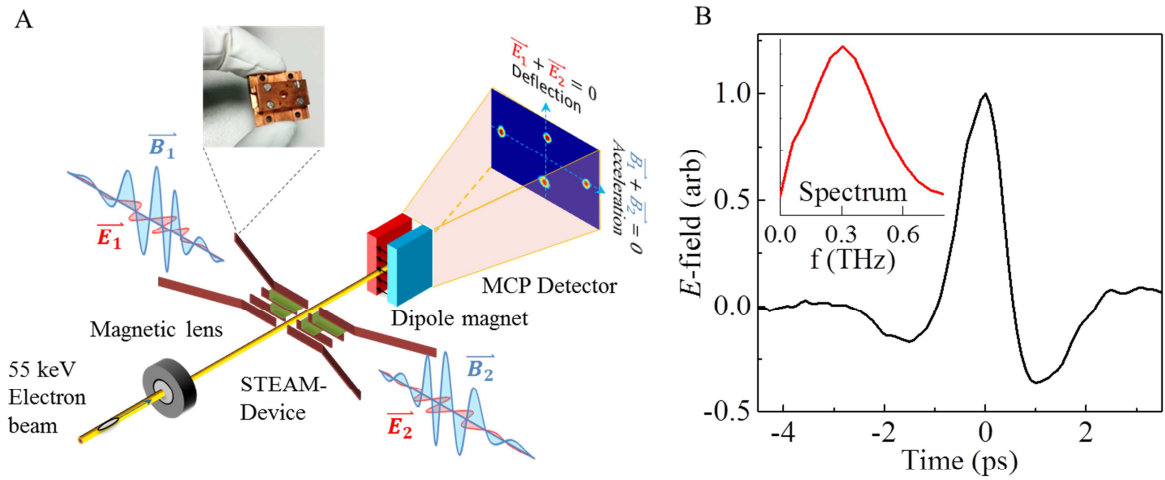

**Fig. 1. Experimental setup.** (A) Schematic illustration of the experimental setup. A fraction of the infrared optical beam generating the THz radiation is converted to 257 nm through fourth harmonic generation. The 257 nm laser pulse is directed onto a gold photocathode generating electron pulses, which are accelerated to 55 keV by a DC electric field. This laser also drives two optical-rectification stages, each generating single-cycle terahertz pulses with energy up to 30 µJ. The two counter-propagating THz beams interact with the 55 keV electron beam inside the segmented THz waveguide structure. Subsequently, the electron beam is detected by the camera. Inset: photograph of the STEAM device. (B) The time-domain waveform of the THz pulse is measured by electro-optic sampling. Inset: corresponding frequency-domain spectrum.

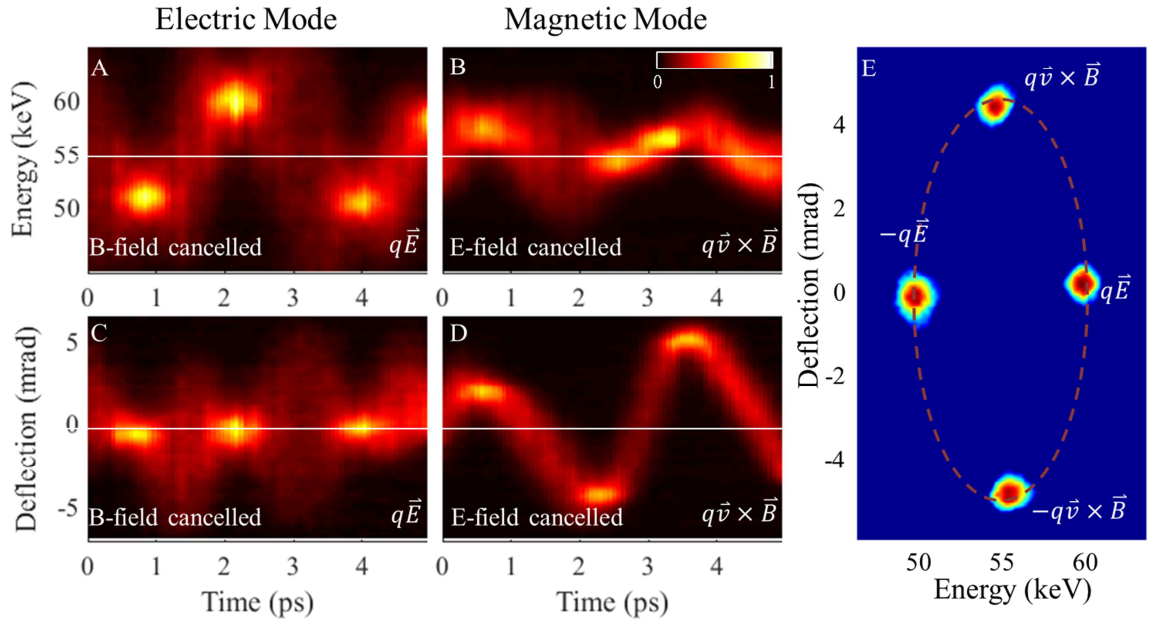

**Fig. 2 Concept and implementation.** (A) Measured energy modulation of e-pulse as a function of delay between THz and e-pulse for constructive interference of the E-fields entering the device and cancelation of *B*-fields. (B) Corresponding beam deflection measured for constructively interfering of the *B*-fields, i.e. *E*-field cancellation scenario. (C) and (D): Time-dependent deflection diagram measured by varying the delay between the e-pulse and the THz-pulse in the *B*-field and *E*-field cancellation scenario, respectively. (E) Measured shape of e-beam on MCP detector for maximum acceleration, deflection and right and left deflection points plotted in one image. Intensity was normalized and image contrast was tuned in order to show the relative positions more clearly. This demonstration was performed with the Yb:KYW laser system with maximum 2×1 µJ of THz radiation coupled into the device.

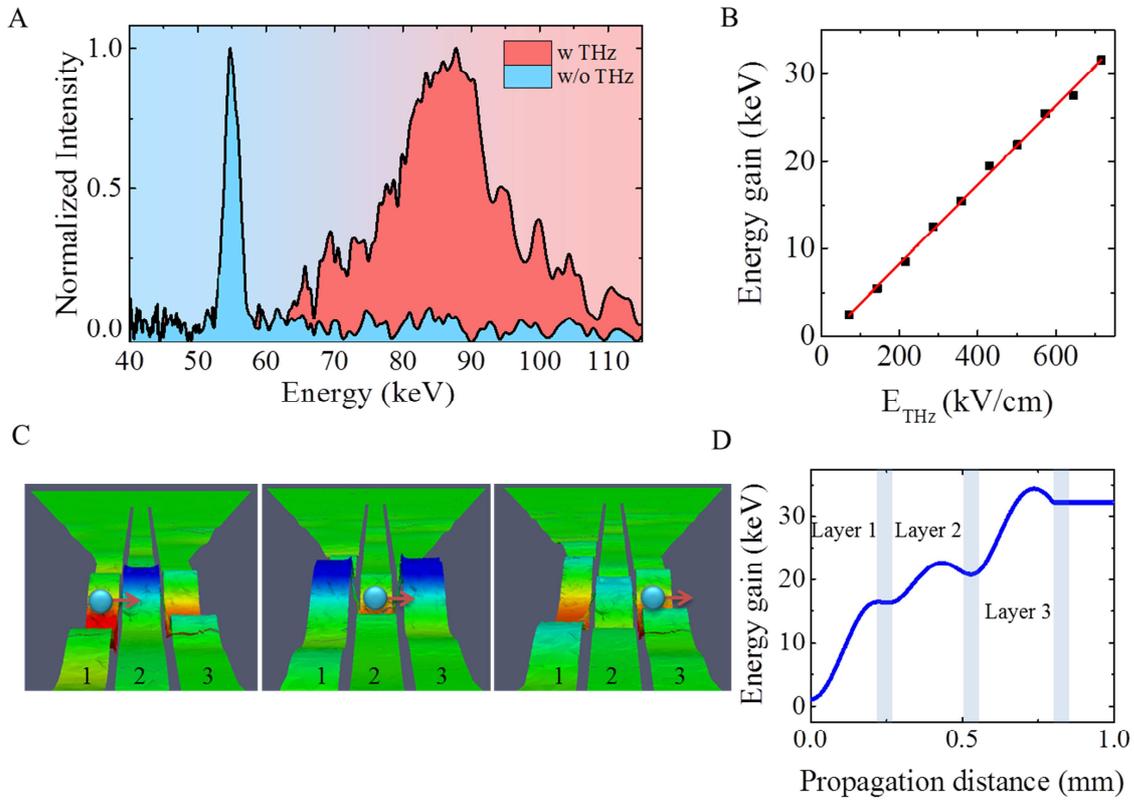

**Fig. 3 Terahertz acceleration.** (A) Measured electron energy spectra for input beam (black curve) and accelerated beam (red curve) after the STEAM-device. An increased energy spread is observed due to the long length of the initial electron bunch, as well as the slippage between the THz pulse and the electron bunch. (B) Relative energy gain versus input terahertz field strength. The linear relationship supports a direct, field-driven interaction. (C) Temporal evolution of the electric field inside each layer with the red arrow indicating the electron propagating direction. (D) Calculated acceleration along the electron propagation direction with ~2×6 μJ THz coupled into the device. This illustration was performed for the experiment using the Yb:YLF laser system with ~2×6 μJ THz coupled into the device.

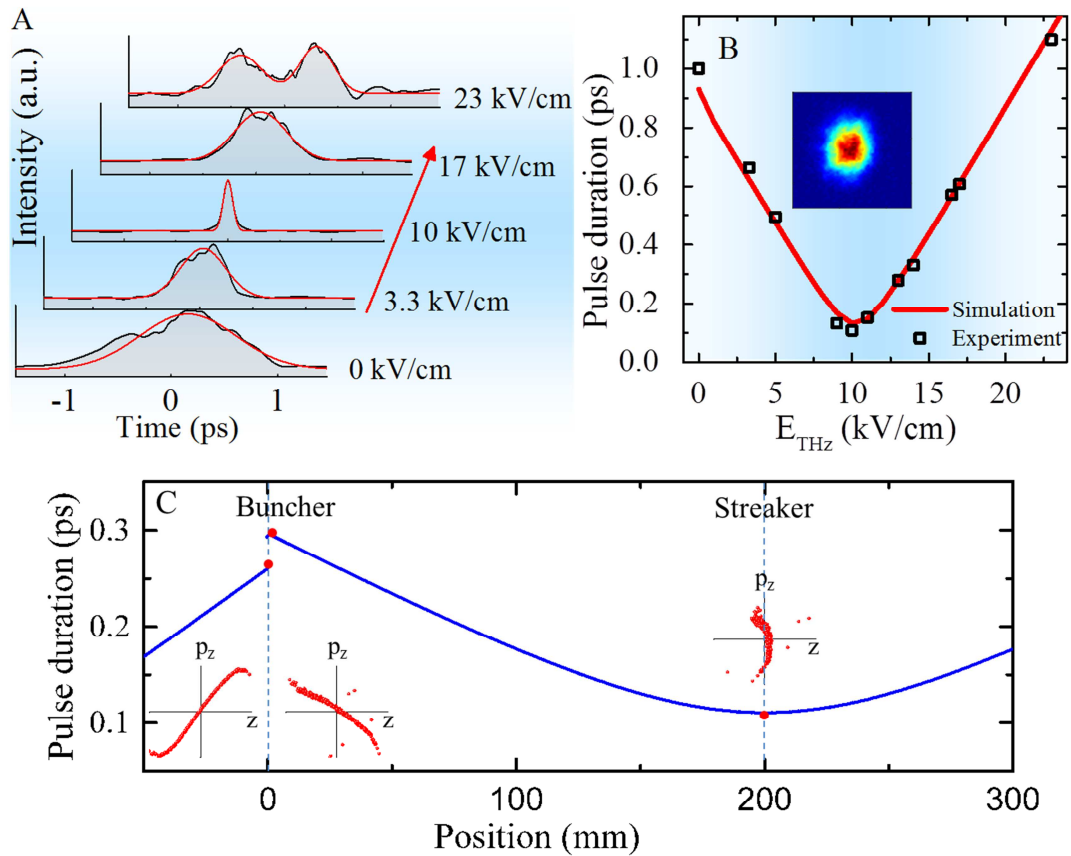

**Fig. 4 Terahertz-driven electron pulse compression.** (A). Measured temporal profiles of the electron pulses as the THz field in the buncher is increased (red arrow). The red lines represent Gaussian fits. (B) Black squares: measured electron bunch FWHM duration versus incident THz field strength. Red line: corresponding simulation results. Inset: the electron beam spatial profile on the detector at the optimal compressed condition. (C) Simulated bunch length vs position. Inset: longitudinal phase-space distribution before the rebunching cavity, after the rebunching cavity and maximally compressed position marked with red dots respectively. This demonstration was performed with the Yb:KYW laser system using one STEAM-device as a rebunching cavity and one as electron streak camera.

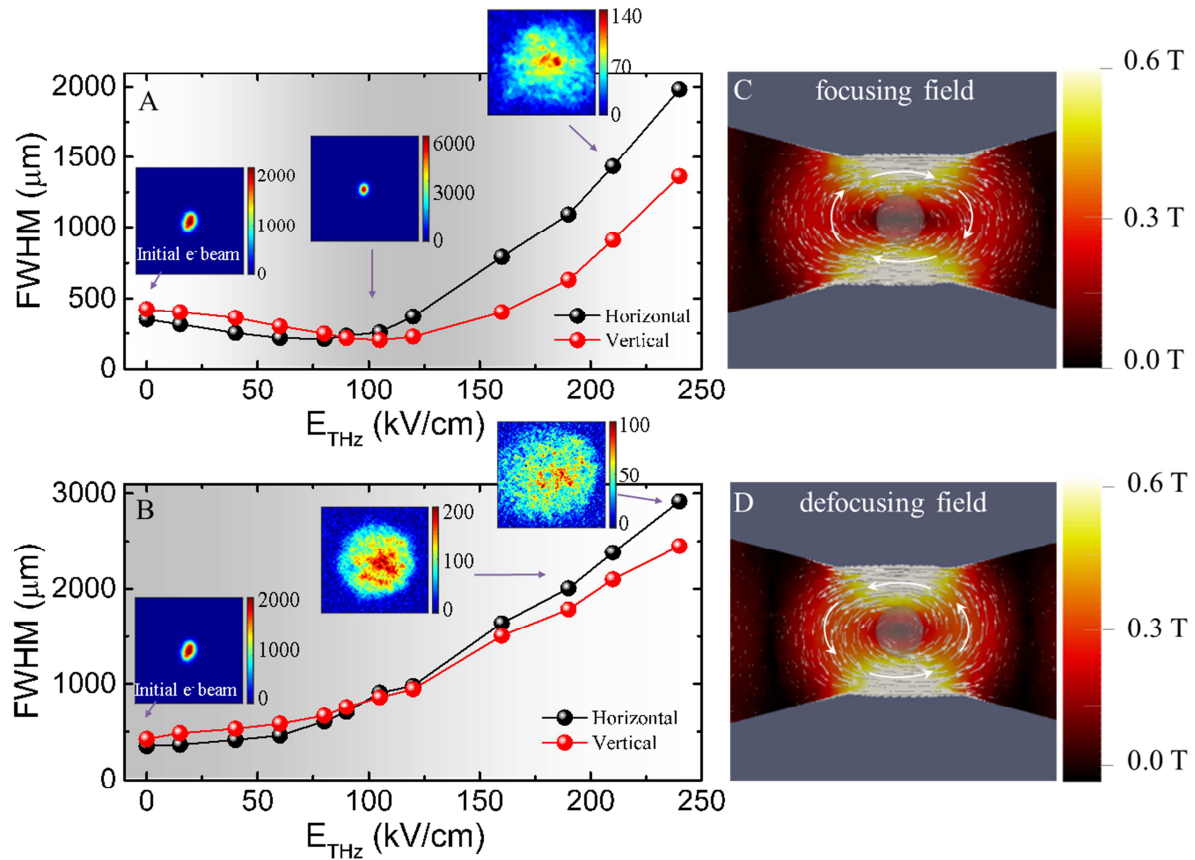

**Fig. 5 Terahertz-lens for electron pulse focusing/defocusing.** (A) and (B): Measured transverse electron beam size at the MCP as a function of THz field, respectively. Insets: spatial profiles of initial and focused and defocused electron beams on the MCP-detector. Legend shows the number of counts on the detector. (C) and (D): Computed spatial field distributions of the focusing and defocusing fields, respectively.

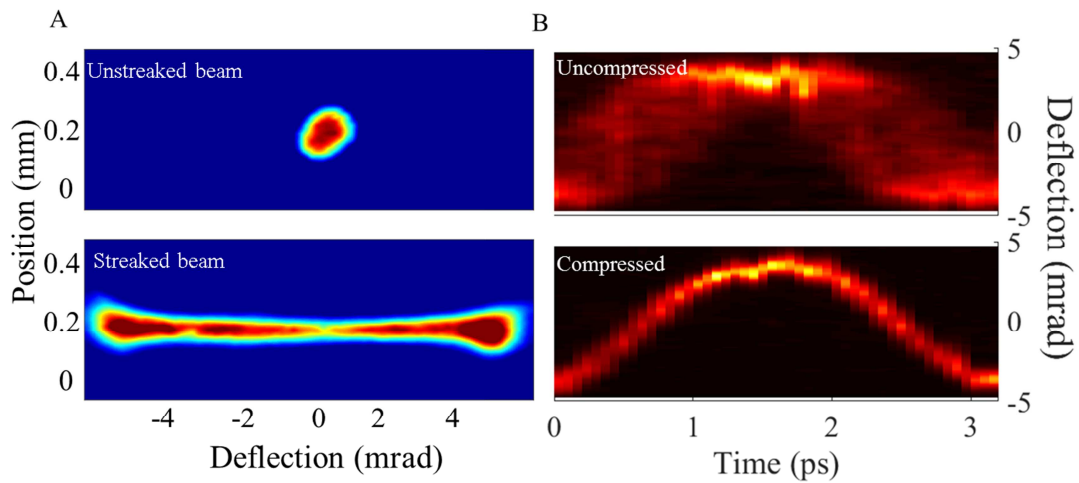

**Fig. 6. Terahertz streak camera.** (A) Measured images of the electron beam on the MCP-detector with and without the terahertz deflection field. (B) Time-dependent deflection diagram measured by varying the delay between the arrival time of electron bunch and the deflecting terahertz pulse for initially compressed and uncompressed electron bunches using the Yb:KYW laser system. Deflectograms using the Yb:YLF laser system can reach more than 70 mrad deflection.